\documentclass[aps,twocolumn]{revtex4}
\usepackage{dcolumn}
\usepackage{graphicx}
\usepackage{amsmath}
\usepackage{amsfonts}
\usepackage{amssymb}
\usepackage{psfrag}
\usepackage{wrapfig}
\usepackage{subfigure}
\usepackage{makeidx}
\usepackage{bm}
\usepackage{epsf}
\usepackage{multirow}
\usepackage[colorlinks,urlcolor=blue,citecolor=blue]{hyperref}

\begin{document}
\title{Semiclassical Trajectory Perspective of Glory Rescattering  in Strong-field Photoelectron Holography}
\author{L. G. Liao$^{1,2}$}
\author{Q. Z. Xia$^{3}$}
\email[]{xia_qinzhi@iapcm.ac.cn}
\author{J. Cai$^{4}$}
\author{J. Liu$^{2,5}$ }
\email[]{jliu@gscaep.ac.cn}
\address{$^{1}$School of Physics, Peking University, Beijing 100871, China}
\address{$^{2}$CAPT, HEDPS, and IFSA Collaborative Innovation Center of MoE, Peking University, Beijing 100871, China}
\address{$^{3}$National Laboratory of Science and Technology on Computational Physics, Institute of Applied Physics and Computational Mathematics, Beijing 100088, China}
\address{$^{4}$School of Physics and Electronic Engineering, Jiangsu Normal University, Xuzhou 221116, China}
\address{$^{5}$Graduate School of China Academy of Engineering Physics, Beijing 100193, China}

\begin{abstract}
We investigate theoretically the photoelectron momentum distribution (PMD) of the ionized atoms irradiated by a linearly polarized intense laser, focusing on the holography interference patterns in PMD that carry important information of the initial wavefunction of a tunneled electron and its experienced atomic potential in rescattering. With the help of Dyson series and semiclassical propagator, we calculate the scattering amplitudes in the cylindrical coordinate representation.
In contrast to conventional recognitions that the photoelectron holography is the interference of two branches of electron trajectories, however,
we find strikingly that infinite semiclassical trajectories can be deflected by the combined Coulomb potential and laser field into the same final momentum: The initial momenta are found to be distributed on a ring-shape curve in the transverse momentum plane and the initial positions of these trajectories are perpendicular to their initial momentum vectors.
For the zero final transverse momentum, the above ring-source trajectories degenerate into the point-source axial caustic trajectories (or Glory trajectories) and the quantum interference of these trajectories will dramatically alter the scattering amplitudes that is termed as Glory rescattering effect.
With following Berry's spirit of uniform approximation for Glory scattering in optics, we can finally derive a uniform formulation of the rescattering amplitude in the Bessel functions for the strong-field photoelectron holography (SFPH) patterns. Our results are in good agreement with solutions of the time-dependent Schr\"odinger equation and can account for recent photoelectron holography experiments. Important applications of our theory are also discussed.
\end{abstract}
\maketitle
\section{INTRODUCTION}
Strong-field photoelectron holography (SFPH) provides a powerful tool for investigating the structure and dynamics of atoms and molecules \cite{PhysRevLett.95.040401,PhysRevLett.103.053001,PhysRevLett.108.223601,PhysRevLett.108.193004,huismans2011time,PhysRevLett.109.013002,walt2017dynamics,haertelt2016probing,meckel2014signatures,PhysRevLett.109.073004,bian2012attosecond,bian2011subcycle,PhysRevA.104.013109}. The physics behind SFPH is that in analog to conventional optical holography, the modulation patterns in photoelectron momentum distribution (PMD) are the phase interference of diverse electron trajectories and therefore carry important informations of initial states of tunneled electrons and their  experienced rescattering potentials \cite{corkum2007attosecond,corkum2011recollision,PhysRevA.102.013109}. Nevertheless, to correctly extract the information contained in the holographic patterns of PMDs, sophisticated nonperturbative theories for the photoionization and rescattering in combined Coulombic and intense laser fields are required.
\par
Semiclassical dynamics can provide intuitive pictures for the strong field ionization and successfully explained the physical mechanisms behind many striking structures in PMD spectroscopy \cite{PhysRevLett.62.1259,PhysRevLett.71.1994,ROST1998271,PhysRevA.62.053403,liu2013classical}. In the semiclassical description, the photoelectron experiences a tunneling through the electromagnetic field suppressed Coulomb barrier and then accelerated in the combined laser field and the Coulomb potential of its parent ion. The latter process is termed as rescattering or recollision. SFPH is tightly related to the quantum interference in the rescattering process \cite{PhysRevA.51.1495,PhysRevA.86.053403,li2015revealing,PhysRevA.100.023413,PhysRevA.100.023419}.
Numerically, PMDs can be obtained by directly integrating the time-dependent Schr\"odinger equation (TDSE) \cite{huismans2011time,PhysRevLett.109.013002,BAUER2006396,PhysRevA.89.023421} in full dimensionality (3D). Theoretically, quantum scattering theory based on strong field approximation (SFA) \cite{keldysh1965diagram,Faisal_1973,PhysRevA.22.1786} as well as the Coulomb corrected strong field approximation (CCSFA) \cite{PhysRevLett.105.253002,popruzhenko2008strong,PhysRevA.77.053409} has been successfully exploited to understand many interesting structures of PMDs.
More recently, Gouy's phase \cite{gouy1890propriete} anomaly or Maslov phase \cite{gutzwiller2013chaos} in strong field rescattering trajectories are addressed \cite{PhysRevLett.124.153202}.
In the above discussions, two branches of trajectories, one is directly ionized trajectory, the other is rescattering trajectory, are considered.
However, due to the existence of caustic singularity \cite{sikivie1999caustic,nussenzveig2006diffraction}, the scattering theories  that coherently sum over two  trajectories for each asymptotic momentum fail to analyze Glory effect, in which the contribution of infinite trajectories can dramatically modulate the scattering amplitude. In order to resolve this caustic singularity, the Glory rescattering theory (GRT) has been developed \cite{PhysRevLett.121.143201}. According to GRT, the infinite semiclassical trajectories are integrated  to give rise to a pattern of Bessel function distribution.
The results of GRT are also certificated by recent two-step model calculations \cite{PhysRevA.100.023419}.
Nevertheless, the GRT focuses on caustic singularity, i.e., the near-zero final transverse momentum region in PMDs. To extend the GRT to non-zero final transverse momentum region and give a uniform description to PMDs from the semiclassical trajectory perspective is urgently needed for practical SFPH applications. \par

On the other hand, M. V. Berry in 1960s has applied a uniform approximation theory to solve the  Glory scattering problems in optics and developed a scattering amplitude formula that can be continually applied from small angle scattering to large angle or even the back scattering \cite{10.2307/43423745,Berry_1969}. Following the concept of the uniform approximation, in the present work, we extend the GRT by exploiting a semiclassical path integral to deduce a uniform formulation  for all angles of forward and backward rescattering in PMDs. Our results are compared with  TDSE calculations, the existing theories as well as the holography experiments.\par

The paper is organized as follows. In Sec. II, we present our theoretical formulation. Sec. III is our results and discussions.\par
Atomic units are used unless otherwise specified.\par

\section{THEORETICAL FORMULISM}
\subsection{Scattering amplitude from the semiclassical trajectory perspective}
We start with the Hamiltonian for atom-field interaction problem in the length gauge, which takes following form of $H[\vec{r}(t),\vec{p}(t)]=\frac{p^2(t)}{2}+\vec{F}(t)\cdot\vec{r}(t)-\frac{1}{r(t)}$.
With exploiting the Dyson series \cite{PhysRev.75.1736} we have, $U(t,0)=U_0(t,0)-i\int_0^tdt_0U(t,t_0)V_L(t_0)U_0(t_0,0).$
Where $U$ denotes the complete evolution operator in the combined Coulomb potential and laser field; $U_0$ represents the evolution operators under pure Coulombic potential, and  has the property of $U_0(t_0,0)|\psi(0)\rangle=|\psi(0)\rangle e^{iI_pt_0}$, here $I_p$ is the ionization potential and $|\psi(0)\rangle$ denotes the initial wavefunction; $V_L=\vec{F}(t)\cdot\vec{r}$ and $\vec{F}(t)$ is the electric field.

With the help of Dyson's series, we thus have time dependent wavefunction $|\psi(t)\rangle=U_0(t,0)|\psi(0)\rangle-i\int_0^t dt_0U(t,t_0)V_L(t_0)U_0(t_0,0)|\psi(0)\rangle.$ The scattering amplitude of a continuum state with asymptotic momentum $\vec{p}_f$ can be written as $M_{\vec{p}_f}=\langle\vec{p}_f|\psi(t_f)\rangle=-i\int dt_0 \langle\vec{p}_f|U(t_f,t_0)V_L(t_0)U_0(t_0,0)|\psi(0)\rangle$ with $t_f \rightarrow \infty$. With inserting an identity $\int d\vec{p}_0|\vec{p}_0\rangle\langle\vec{p}_0|\equiv 1$, we have

\begin{eqnarray}
M_{\vec{p}_f}=-i\iint dt_0d\vec{p}_0
G(\vec{p}_f,t_f;\vec{p}_0,t_0)\mathcal{D}(\vec{p}_0, t_0)e^{iI_pt_0}.
\label{eqn1_2}
\end{eqnarray}
Here $\mathcal{D}(\vec{p}_0, t_0)=\langle \vec{p}_0| V_L(t_0)|\psi(0)\rangle$ represents the dipole matrix element and the momentum-to-momentum propagator $G(\vec{p}_f,t_f;\vec{p}_0,t_0)=\langle \vec{p}_f|U(t_f,t_0)|\vec{p}_0\rangle$.

In the semiclassical approximation \cite{gutzwiller1967phase,gutzwiller2013chaos}, the propagator $G(\vec{p}_f,t_f;\vec{p}_0,t_0)=\mathcal{F}(\vec{p}_f,t_f;\vec{p}_0,t_0) e^{i\mathcal{S}(\vec{p}_f,t_f;\vec{p}_0,t_0)}$,  where the semiclassical phase $\mathcal{S}(\vec{p}_f,t_f;\vec{p}_0,t_0)=\int_{t_0}^{t_f}dt(-\vec{r}(t)\cdot\dot{\vec{p}}(t)-H[\vec{r}(t),\vec{p}(t)])$. We neglect the Maslov phase in holography patterns, and the reason is discussed in subsection C. The prefactor $\mathcal{F}=[\frac{1}{(2\pi i)^3}\det(\frac{\partial^2 \mathcal{S}}{\partial p_0\partial p_f})]^{\frac{1}{2}}$.

Considering the case of linearized laser field that is polarized along z axis, the problem intrinsically has the rotational symmetry about the axial coordinate. We therefore introduce the cylindrical coordinate instead of Cartesian coordinate. Thus, Eq. (\ref{eqn1_2}) can be rewritten as
\begin{equation}
  M_{\vec{p}_f}=\iiiint dt_0dp_{z0}dp_{\rho 0}d\phi p_{\rho 0}
\mathcal{F}\cdot \mathcal{D}\cdot
  e^{{i\mathcal{A}(\vec{p}_f,t_f;\vec{p}_0,t_0)}}
  \label{cylindrical}
\end{equation}
Here, $\mathcal{A}=\mathcal{S}+I_pt_0$.
Because $\mathcal{F}$ and $\mathcal{D}$ are slowly-varying functions, we treat Eq. (\ref{cylindrical}) with the steepest descent method for the time integration as well as the momentum integrations. The saddle point condition for the time variable gives,
\begin{eqnarray}
\frac{\partial \mathcal{A}}{\partial t_0}|_{t_s}=0.
\label{time_saddle}
\end{eqnarray}
With applying the steepest descent method to treat the integration on $p_{z0}$ and $p_{\rho 0}$, we obtain the  coordinates of the saddle points satisfying
\begin{eqnarray}
&z(t_{s})=0 ; \,\,\,\,\,\,\vec{\rho}_s\cdot\vec{p}_{\rho s}=0.
\label{eq9_1}
\end{eqnarray}
In the above deductions, we use the relation
$\frac{\partial \mathcal{A}}{\partial p_\rho}=x\cos\phi+y\sin\phi=\vec{\rho}\cdot\vec{p_\rho}/|p_\rho|$.
In contrast to conventional steepest descent method implemented in the Cartisian coordinates, in which the saddle point conditions give that the positions of the saddle points are located at origin, in our cylindrical coordinate representation, the saddle points might even not be on the symmetric z-axis. The second formula in  Eq. (\ref{eq9_1}) gives a constraint on the initial coordinates and momenta. \par
Then the scattering amplitude here becomes
\begin{eqnarray}
  M_{\vec{p}_f}&\approx \int d\phi p_{\rho_{s}}(\frac{(2\pi i)^3}{\det(\frac{\partial^2\mathcal{A}}{\partial(t_{0},{p}_{z0},p_{\rho 0})}|_s)})^{\frac{1}{2}}\mathcal{F}_s\mathcal{D}_se^{i\mathcal{A}_s}.
\label{eq_phi}
\end{eqnarray}
where $\mathcal{A}_s$,$\mathcal{F}_s$, and $\mathcal{D}_s$ represent the values at the saddle point $(\vec{p}_{s},t_{s})$,\par

In the above integration, the phase $\mathcal{A}_s$ is the implicit function of the azimuthal angle $\phi$.  To treat this scattering amplitude integration, we closely follow the Berry's spirit of uniform approximation \cite{10.2307/43423745,Berry_1969} with renormalizing the angular variable. Let us introduce a new angular variable $\varphi$, which satisfies $\phi=0\leftrightarrow\varphi=0$, $\phi=\pi\leftrightarrow\varphi=\pi$ and $\mathcal{A}_s(\phi)=\bar{\mathcal{A}}_s(\varphi)\equiv\mathcal{A}_0+\mathcal{A}_1\cos(\varphi)$, where $\mathcal{A}_0=(\mathcal{A}_s(\phi=0)+\mathcal{A}_s(\phi=\pi))/2$ corresponds to the sum of the phases of the $\phi=0$ and $\phi=\pi$ semiclassical photoelectron trajectories, $\mathcal{A}_1=(\mathcal{A}_s(\phi=0)-\mathcal{A}_s(\phi=\pi))/2$ corresponds to the phase difference between the $\phi=0$ and $\phi=\pi$ trajectories.

With the variable transformation, the scattering amplitude of Eq. (\ref{eq_phi}) turns to be (see Appendix A for details), \par
\begin{equation}
\begin{split}
M_{\vec{p}_f}\approx \int_0^{2\pi}d\varphi C(\varphi)
e^{i(\mathcal{A}_0+\mathcal{A}_1\cos\varphi)}.
\label{eq_cphi}
\end{split}
\end{equation}
 The prefactor $C(\varphi)$ takes form,
\begin{eqnarray}
C(\varphi)\propto \frac{\mathcal{D}_s}{\sqrt{\frac{d^2\mathcal{A}_s}{dt_{s}^2}}}(\mathcal{A}_1\cos\varphi)^{\frac{1}{2}}\frac{p_{\rho s}}{p_{\rho f}}\det(\frac{\partial (p_{zs},p_{\rho s})}{\partial(p_{zf},p_{\rho f})}).
\label{eq7}
\end{eqnarray}
We can approximate the the above function in the following form $C(\varphi)\approx \frac{1}{2}(C(0)+C(\pi))+\frac{1}{2}(C(0)-C(\pi))\cos\varphi$. Finally, the scattering amplitude integration can be evaluated by the Bessel functions in following explicit forms, that is termed as uniform Glory rescattering theory (UGRT).
\begin{equation}
\begin{split}
&M_{\vec{p}_f}\\
\approx&\frac{1}{2}\int_0^{2\pi}d\varphi[(C(0)+C(\pi))+(C(0)-C(\pi))\cos\varphi]e^{i\mathcal{A}_1\cos\varphi}\\
=&\frac{1}{2}(C(0)+C(\pi))J_0(\mathcal{A}_1)-\frac{1}{2}i(C(0)-C(\pi))J_1(\mathcal{A}_1)
\end{split}
\label{final}
\end{equation}

Let us now analyze the physics behind the above deductions.
From the expression of Eq. (\ref{eq_phi}), we see that for a given final photoelectron momentum, in contrast to conventional recognition of the two-trajectory interference, we find that infinite semiclassical trajectories can be  deflected by the combined Coulomb potential and laser field into the same final momentum. This phenomenon is  mentioned very recently only for the case of zero final transverse momentum \cite{PhysRevLett.121.143201}, where
the infinite trajectories of tunneling electrons at moment $Re(t_s)$ on z-axis  with equal transverse momentum are rescattered to the final states with zero transverse momentum, leading to so called caustic singularity\cite{doi:10.1080/00107514.2015.971625}. Similar phenomenon is termed as the Glory scattering first discussed in optics  and then extended to particle scattering by Wheeler \cite{ford1959semiclassical,nussenzveig2006diffraction,ADAM2002229}. The quantum interference of the infinite Glory trajectories will manipulate the scattering amplitude with showing a bright fringe around the zero angle in SFPH. This picture is apparent and can be readily imagined by considering the cylindrical symmetry of the geometric configuration of the problem.

Our above deductions explicitly indicate that the infinite-trajectory interference can emerge also for nonzero final transverse momenta.  According to Eq. (\ref{eq9_1}) and from our detailed  calculations of these semiclassical trajectories (see Fig. \ref{fig1} and \ref{fig2} in following section), we find that the initial position of these trajectories  are no longer on the symmetric z-axis and found to be distributed on a ring-shape curve in the transverse coordinate plane.  The initial transverse momentum are perpendicular to the initial position vectors according the constraint  of Eq. (\ref{eq9_1}).

Moreover, the expression of Eq. (\ref{final}) shows that the scattering amplitude  can be expressed in the sum of zeroth-order and first-order Bessel functions in uniform approximation.
The variable $\mathcal{A}_1$ in the Bessel functions represents the phase difference of two distinct semiclassical trajectories corresponding to $\phi=0$ and $\phi=\pi$, respectively.
When the final transverse momentum tends to 0, $C(0)$ equals to $C(\pi)$, and Eq. (\ref{final}) will reduce to the formulation of GRT in Ref. \cite{PhysRevLett.121.143201}.
In GRT,  $M_{\vec p}$ can be written in the following form $|M_{\vec p}|^2\sim\varpi p_{\perp g}b_gJ_0^2(p_{\perp}b_g)$ in which $p_{\perp g}$ is the initial transverse momentum of the Glory trajectory at the tunneling exit, $b_g$ is the emergent impact parameter of the Glory trajectory, and $\varpi$ is the weight of the Glory trajectory.
In this situation,  we note that the phase difference of  $\mathcal{A}_1$ in Eq. (\ref{final}) approximately equals to $p_{\perp}b_g$, here $p_{\perp}$ is the final transverse momentum.

\subsection{Calculations of semiclassical trajectories and associated phase accumulations}
\begin{table}[b]
\caption{\label{tab:table1}States of an electron in ionization process}
\begin{ruledtabular}
\begin{tabular}{ccc}
 Time&State description&Position and/or momentum
\\ \hline
 $t_{s}$&saddle point&$z_{s}=0, \vec{\rho}_s\cdot\vec{p}_{\rho s}=0,\vec{p}_{z s},\vec{p}_{\rho s}$  \\
 $Re(t_s)$&tunneling exit
 &$\vec{r}_0=Re\int_{t_{s}}^{Re(t_s)}(\tilde{\vec{p}}+A(t))dt,\vec{p}_{s}$\\
 $t_f=\infty$&final state&$\vec{p}_f, \dot{\vec{p}}=-\vec{F}(t)-\vec{\nabla}V(\vec{r(t)})$
\end{tabular}
\end{ruledtabular}
\end{table}
States at some typical times that an electron experiences in its ionization process  are listed in TABLE \ref{tab:table1}, in which $Re(t_s)$ is the real part of saddle point time $t_{s}$. The condition of saddle point according to Eq. (\ref{time_saddle}) is obtained by solving saddle point equation within the strong-field approximation  (see, e.g., \cite{Milo_evi__2006,PhysRevA.93.013402}) $\frac{1}{2}({{\tilde p}_{\rho s} ^2+[\tilde p_{zs}+A(t_s)]^2})+I_p=0$, here $\tilde {p}$ is the canonical momentum. The corresponding kinetic momenta at saddle point are then given by $p_{zs}=\tilde {p}_{zs}+A(t_s)$ and $p_{\rho s}=\tilde {p}_{\rho s}.$

Under the constraint of saddle point condition of Eq. (\ref{eq9_1}), for a given final momentum, we can use shooting method \cite{morrison1962multiple} to obtain the initial conditions of semiclassical trajectories, by solving following Newton equations of motion numerically using the Runge-Kutta-Fehlberg method, $\dot{\vec{p}}=-\vec{F}(t)-\vec{\nabla}V(\vec{r}(t))$ where $V(\vec{r}(t))$ is the atomic potential. The asymptotic condition is that $t\rightarrow\infty$, $\vec{p}\rightarrow\vec{p}_f$. In semiclassical treatment, the phase of each trajectory is divided into two parts: ionization process phase and acceleration process phase.
The phase accumulated during ionization process is given by $\mathcal{A}(t_{s}\rightarrow Re(t_s))=-\frac{1}{2}\int_{t_{s}}^{Re(t_s)}({{\tilde p}_{\rho s} ^2+[\tilde p_{zs}+A(t)]^2})dt.$
The  phase accumulated during the acceleration process can be expressed as $\mathcal{A}(t_f;Re(t_s))=- \int_{Re(t_s)}^{t_f}dt\{\dot{\vec{p}}(t)\cdot\vec{r}(t)+H[\vec{r}(t),\vec{p}(t)]\}$ \cite{PhysRevA.94.013415}.

With these semiclassical trajectories and corresponding  phases, we can reconstruct the photoelectron momentum spectrum  according to Eq. (\ref{final}).

\subsection{Some remarks}
In conventional treatments, one directly treat integral of Eq. (\ref{eqn1_2}) with steepest descent method for all momentum variables in Cartesian coordinates, the initial positions of tunneled electron will locate on z-axis. Then, within half a laser cycle there are usually two trajectories (one is directly ionized trajectory, the other is rescattering trajectory) can reach the same final momentum. The  two-trajectory interference is used to explain for the SFPH where the modulation fringe can be expressed in the cosine function of the phase difference between the two trajectories.
Recently, the Gouy's phase \cite{gouy1890propriete} is introduced to the two-trajectory strong-field interference picture in a 3D model to compensate for the divergence of the preexponential factors of the semiclassical propagator at focal points \cite{PhysRevLett.124.153202}. As a result, phase difference in the cosine function will be modified by a $\nu\pi/2$ phase, where $\nu$ is the Maslov index \cite{maslov2001semi,mohring1980semiclassical,du1987effect,delos2009semiclassical}.
In fact, due to the cylindrical symmetry of the problem, infinite ring-source trajectories can converge to the same final momentum state, the quantum interference of the ring-source trajectories will give rise to the pattern in PMDs of Bessel functions instead of cosine functions.
Within the framework of UGRT, the Maslov phases of different trajectories vanish.
This is because in the representation of cylindrical coordinates, due to the cylindrical symmetry, the system essentially reduce to a 2D problem under the uniform approximation, in which the preexponential factors of the semiclassical propagator keep finite along the rescattering trajectories (i.e., no focal points)  according to our detailed calculations.
\par

\section{APPLICATIONS}
\subsection{Hydrogen atom in linearly laser fields}

\begin{figure}[t]
\includegraphics[width=1.0\linewidth]{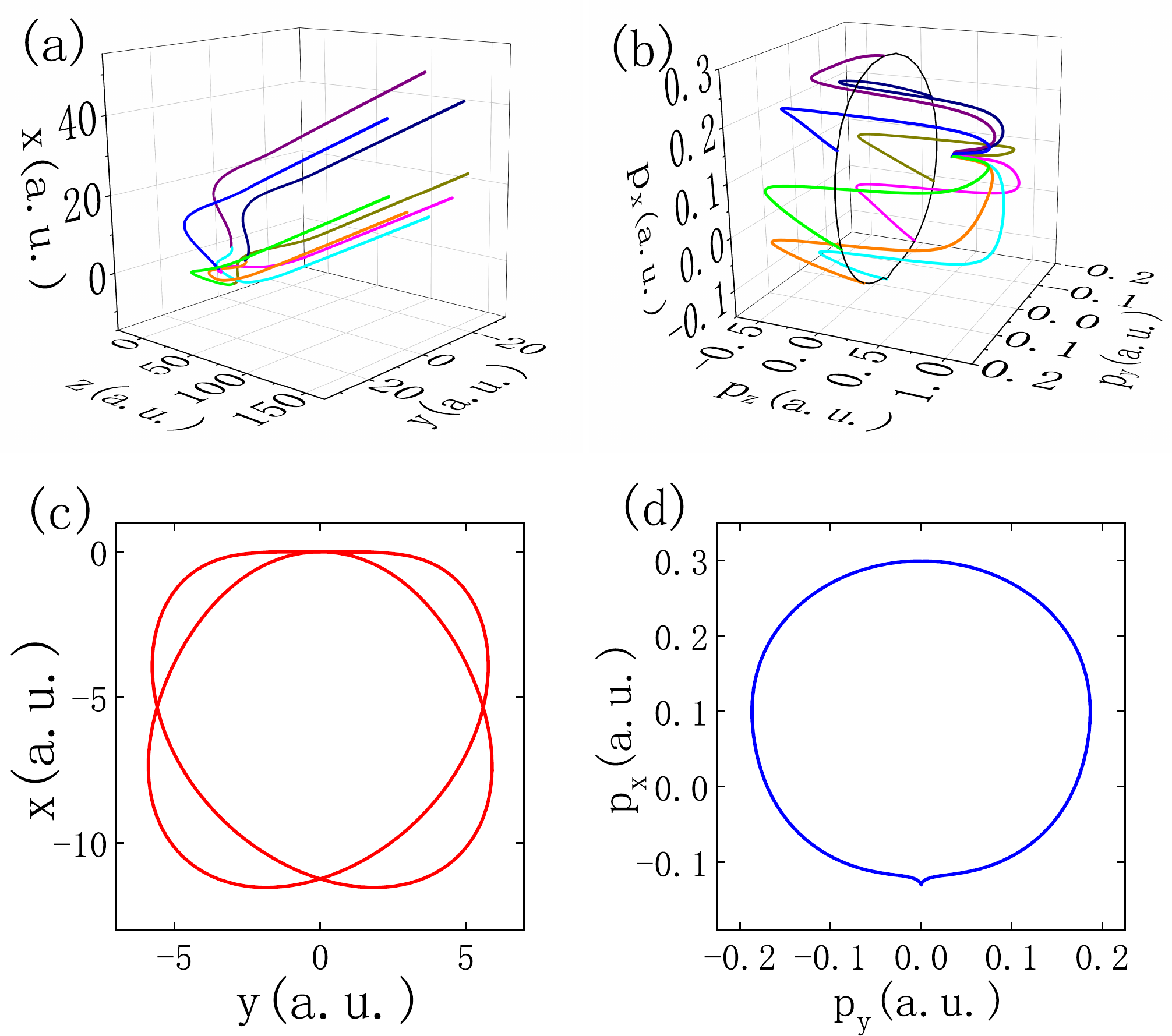}
\caption{\label{}Photoelectron semiclassical trajectories in coordinate space (a) and momentum space (b) for the final state of  $p_{zf}=0.66 a.u.$, $p_{x f}=0.12 a.u.$, $p_{y f}=0$. The initial positions on the projected x-y plane (c) and initial transverse momenta on the projected $p_x$-$p_y$ plane(d). The laser parameters are $\lambda=1200nm$ and $I=8.7\times10^{13}W/cm^2$.}
\label{fig1}
\end{figure}

As a demonstration, we consider a hydrogen atom irradiated by  a few-cycle linearly polarized infrared laser pulse. The vector potential of laser field is
\begin{equation}
  A=-\frac{A_0}{\omega}\sin^2(\frac{\pi t}{t_p})\sin(\omega t)\vec{e}_z.
\end{equation}
The laser field is present between $t=0$ and $t=t_p$ and $\vec{e}_z$ is the unit vector pointing in polarization direction. The electric field is obtained by $\vec{F}=-\frac{1}{c}\frac{\partial A}{\partial(t)}$.

\subsubsection{Ring-source semiclassical trajectories}
We use shooting method to obtain semiclassical trajectories for a given final momentum under the constraint of saddle point conditions given by Eq. (\ref{eq9_1}). Our numerical results show that there are infinite semiclassical trajectories deflected by the combined Coulomb potential and laser field into the same final momentum.
In Fig. \ref{fig1} (a) and (b), we draw some typical trajectories of the tunneling electron in coordinate space as well as  momentum space, and all of these trajectories reach the same final momentum of $p_{zf}=0.66$, $p_{x f}=0.12$ and $p_{y f}=0$ when $t_f \rightarrow \infty$. From Fig. \ref{fig1} (c, d), we see interestingly that the initial momenta of these semiclassical trajectories are distributed in a ring, and the initial coordinates are no longer on the z axis and show a symmetric double-ring structure. After the acceleration of the laser field and the scattering of the nuclear Coulomb potential, these photoelectrons are finally scattered  into the same final momentum (see Fig. \ref{fig1} (b)).

\begin{figure}[t]
\includegraphics[width=1.0\linewidth]{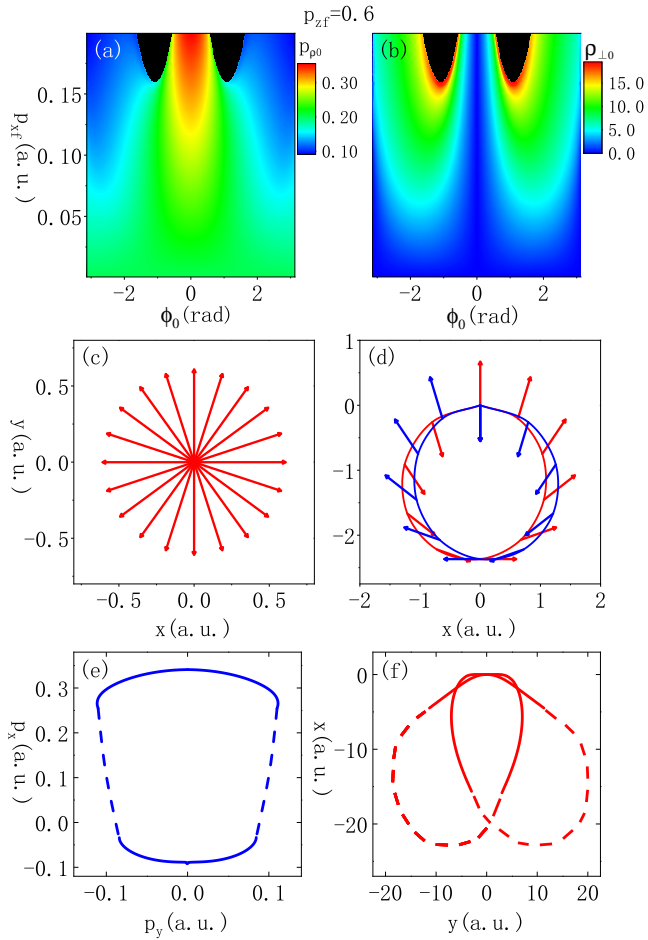}
\caption{\label{}For $p_{zf}=0.6$ a.u., $p_{yf}=0$ a.u. and $p_{xf}$ varied from 0 to 0.2 a.u., we present the calculated initial transverse momentum of $p_{\rho 0}$ depending on azimuthal angle of $\phi_0$ (a), and initial displacement of $\rho_0$ to z-axis depending on azimuthal angle of $\phi_0 $ (b). In (c), for $p_{x f}$=0 corresponding to the Glory caustic singularity, the initial positions are exactly at origin on the projected x-y plane, the initial transverse momenta are schematically by red arrows. In (d), for nonzero final transverse momentum of $p_{xf}= 0.03$ a.u., we plot the initial positions on the projected x-y plane and the initial transverse momenta schematically by red arrows. The initial transverse momenta of photoelectron trajectories when $p_{xf}= 0.2$ a.u. (e) and the corresponding initial positions on the projected x-y plane (f).
The black areas in (a) and (b) represent the classical trajectory forbidden regions. The dashed lines in (e) and (f) is the analytical extension of the solutions to the forbidden regions.
The laser parameters are $\lambda=1200nm$ and $I=8.7\times10^{13}W/cm^2$.}
\label{fig2}
\end{figure}

Fig. \ref{fig2} (a) and (b) shows the initial transverse momentum $p_{\rho 0}$, position $\rho_0$ and azimuthal angle $\phi_0$ with respect to final transverse momentum $p_{x f}$ for the fixed final momentum $p_{zf}$=0.6, $p_{yf}$=0. When final transverse momenta $p_{xf}$ tends to 0, the initial transverse momentum becomes independent on  $\phi_0$, and the initial positions $\rho_0$ tends to zero. This corresponds to Glory rescattering trajectory as shown in Fig. \ref{fig2} (c). For small $p_{xf}$ of 0.03, in Fig. \ref{fig2} (d), we also draw the joint distribution of the initial positions and initial momenta of photoelectron trajectories, which forms ring source similar to that of  Fig. \ref{fig1}.

Interestingly, with increasing the transverse final momentum more ($>0.16$), our calculations show some classical trajectory forbidden regions denoted by the  black areas in Fig. \ref{fig2} (a) and (b). For instance, we plot, for the case of $p_{xf}= 0.2$, the initial transverse momenta of photoelectron trajectories in Fig. \ref{fig2} (e) and the initial positions on the projected x-y plane Fig. \ref{fig2} (f). In our calculations, we confine our initial positions of tunneling electrons within 16 a.u. distance to z-axis.
To apply our UGRT to these situations, we need to make the analytical extension of the solutions to the forbidden regions, as indicated by the dashed lines in Fig. \ref{fig2} (e) and (f).

\subsubsection{Calculations of PMDs}
\begin{figure}[t]
\includegraphics[width=1.0\linewidth]{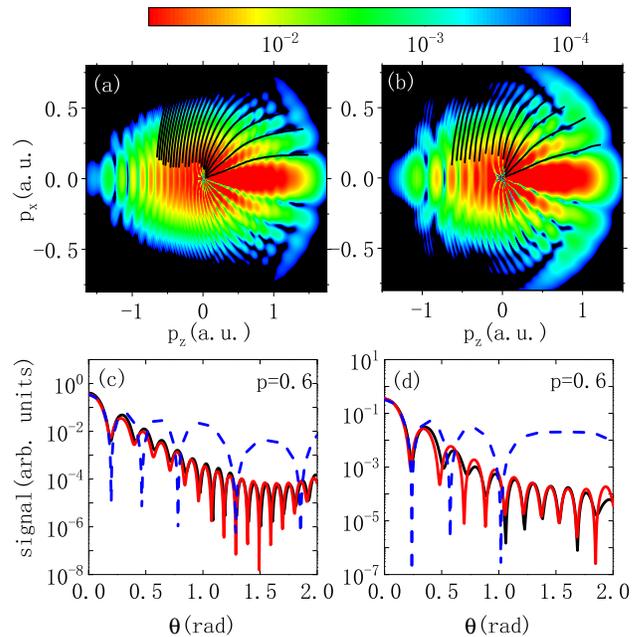}
\caption{\label{} In panels of (a) and (b), we show the PMDs for the hydrogen ionized by laser pulses with $8.7\times10^{13}W/cm^2$ intensity calculated by TDSE, and the interference fringes predicted by our UGRT (black lines) .
In panels of (c) and (d), we plot the momentum distributions corresponding to fixed final momenta $p$=0.6. from calculations of TDSE (black lines), UGRT (red lines), and GRT (blue dash), respectively. The wavelength of laser pulses are $\lambda$ = 1600nm [(a), (c)] and $\lambda$ = 1200nm [(b), (d)]. In (c) and (d), scattering angle $\theta=\arctan(\frac{p_{\rho f}}{p_{z f}})$.}
\label{fig3}
\end{figure}
To validate our theory, we also solve the time-dependent Schr\"odinger equation (TDSE) of a hydrogen (H) atom in infrared few-cycle linearly polarized laser field with a generalized pseudo-spectral method \cite{tong1997theoretical}. The corresponding Hamiltonian is $H[\vec{r}(t),\vec{p}(t)]=\frac{p^2(t)}{2}+\vec{F}(t)\cdot\vec{r}(t)-\frac{1}{r(t)}$.

In Fig. \ref{fig3}, we compare our theoretical results with TDSE for varied  laser wavelengths. From Fig. \ref{fig3} (a) and (b) we can clearly see that the fringes calculated using UGRT is highly consistent with the results of TDSE. UGRT can precisely predict the positions of interference fringes even for the large scattering angles. The 1D slices of PMDs at the fixed final momenta $p$ in Fig. \ref{fig3} (c) and (d) show that the results of UGRT can also quantitatively predict the scattering amplitudes. As a comparison, the GRT can only give a good prediction inside the first scattering minimum but fails  to predict both fringe positions and scattering amplitudes for the higher-order fringes.\par

\begin{table*}[t]
\caption{Optical analogs of SFPH theories ($\Delta$S denotes the phase difference of semiclassical trajectories)}
\begin{ruledtabular}
\begin{tabular}{ccccc}
 Theory&Interference picture&Interference  formula&Optical analogs\\
 \hline
 conventional models&two-trajectory interference & $\cos$($\Delta$S) &double slit interference\\
 UGRT&ring-source infinite trajectories
 &$C_0 J_0(\Delta S)+C_1 J_1(\Delta S)$&ring-source diffraction\\
 GRT&point-source infinite trajectories&$J_0(\Delta S)$&point-source diffraction
\end{tabular}
\end{ruledtabular}
\end{table*}

For the small scattering angle, the scattering amplitude predicted by the two-trajectory interference models (even with Gouy's phase modification) diverges because prefactor $\det(\frac{\partial \vec{p}_{s}}{\partial\vec{p}_f})$ there tends to infinite \cite{PhysRevLett.124.153202}. In our UGRT and GRT, the presence of the term of the square root of the phase difference in the prefactor (see Eq. (\ref{eq7})) will eliminate this singularity. When the final transverse momentum is large, i.e., for the large scattering angles, the phase differences increase. The Bessel functions then reduce to the simple cosine functions in the asymptotical forms of  $J_\alpha(x)\rightarrow\sqrt{\frac{2}{\pi x}}cos(x-\frac{\alpha \pi}{2}-\frac{\pi}{4})$, $\alpha$ is the order of the Bessel function.\par

\subsection{Application to experiment of Xeon}
\begin{figure}
\includegraphics[width=0.55\linewidth]{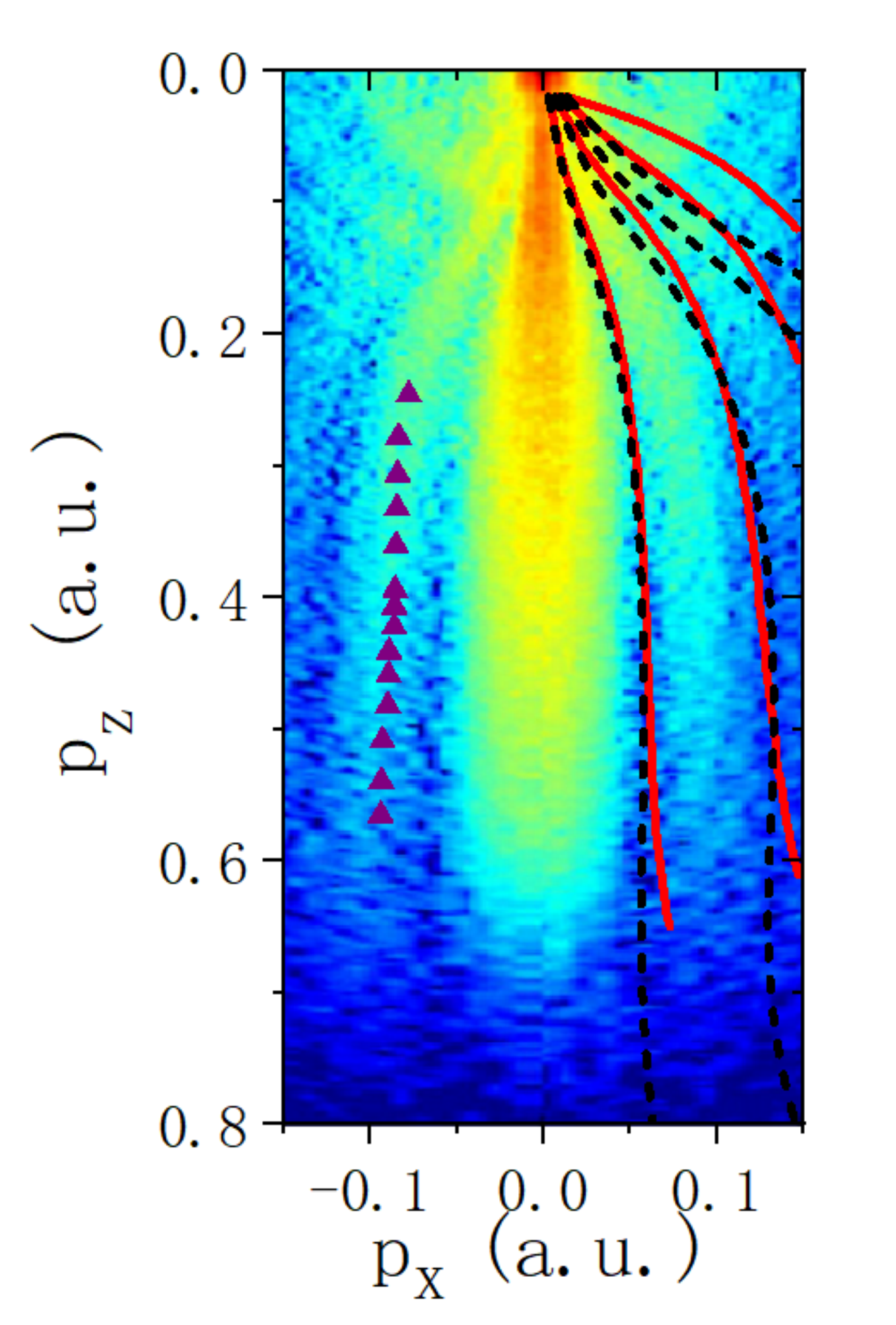}
\caption{\label{}Experimental holographic pattern and positions of the dark fringes calculated by UGRT (red lines), GRT (black dash) and CCSFA (purple triangles). The experimental data is extracted from Ref \cite{PhysRevLett.109.013002}, where the metastable (6s) Xe atoms with ionization potential 0.14 a.u. are ionized with linearly polarized mid-infrared laser. The laser parameters are $\lambda=7\mu m$ and $I=7.1\times10^{11}W/cm^2$.}
\label{fig4}
\end{figure}
We now apply our theory to the experiment of Xeon (the red lines in Fig. \ref{fig4}). The experiments of metastable (6s) Xe atoms in this figure is using 7000 nm laser field \cite{PhysRevLett.109.013002}. As a comparison, we also plot the results predicted by other theories such as GRT (black dash) and CCSFA  (purple solid triangles). According to CCSFA, the coherent summation of the two paths leads to $\cos\Delta S$-type oscillations in holography patterns. We can see that CCSFA even fails to predict the position of the first dark fringe, while GRT and UGRT are in perfect agreement with the experiment in this region. However, the discrepancy between the GRT and UGRT becomes more apparent for the third or higher order fringes. We hope that these theoretical predictions can be calibrated in future by SFPH experiments with higher resolution.

\section{CONCLUSIONS}
We investigate the PMDs of the ionized atoms irradiated by a linearly polarized strong laser field and provide a semiclassical trajectory perspective of Glory rescattering in SFPH. We calculate the scattering amplitudes in the cylindrical coordinate representation and finally derive a uniform formulation in the Bessel functions for the SFPH patterns.
Our results are also in good agreement with solutions of the TDSE calculations and can give explanations to recent photoelectron holography experiments of Xe atoms. Compared with existing theories (see Table II), our UGRT provides a distinct ring-source infinite-trajectory interference description for photoelectron holograph patterns analogous to ring-source diffraction in optics.  Our results of uniform approximation can also be applied to molecule and nondipole situations, related works are undergoing. Our work therefore have important implications in both theoretical aspects and the practical applications of SFPH.

\section*{ACKNOWLEDGMENTS}
This work is supported by the National Natural Science
Foundation of China (Grants No. 11775030 and No.
11974057) and NSAF (Grant No. U1930403).

\appendix

\section{Uniform approximation}
From Eq. (\ref{eq_phi}) we can see that the phase $\mathcal{A}_s$ is the implicit function of the azimuthal angle $\phi$. To treat this scattering amplitude integration, we closely follow the Berry's spirit of uniform approximation \cite{10.2307/43423745,Berry_1969} with renormalizing the angular variable. Let us introduce a new angular variable $\varphi$, which satisfies $\phi=0\leftrightarrow\varphi=0$, $\phi=\pi\leftrightarrow\varphi=\pi$ and $\mathcal{A}_s(\phi)=\bar{\mathcal{A}}_s(\varphi)\equiv\mathcal{A}_0+\mathcal{A}_1\cos(\varphi)$. Where $\mathcal{A}_0=(\mathcal{A}_s(\phi=0)+\mathcal{A}_s(\phi=\pi))/2$ corresponds to the sum of the phases of the $\phi=0$ and $\phi=\pi$ semiclassical photoelectron trajectories, $\mathcal{A}_1=(\mathcal{A}_s(\phi=0)-\mathcal{A}_s(\phi=\pi))/2$ corresponds to the phase difference between the $\phi=0$ and $\phi=\pi$ trajectories. Further, we can get
\begin{equation}
  \frac{d\phi}{d\varphi}=\frac{d\bar{\mathcal{A}}_s}{d\varphi}/\frac{d\mathcal{A}_s}{d\phi}=\frac{\mathcal{A}_1sin\varphi}{L_{\phi}}.
  \label{lhos}
\end{equation}
Where $L_{\phi}=-\frac{d\mathcal{A}}{d\phi}$ is the photoelectron angular momentum along the polarization direction.
For linearly polarized laser field, the mirror symmetry will lead to $\frac{d\mathcal{A}_s}{d\phi}=0$ at $\phi=0$ and $\pi$. Both the denominator and numerator in  Eq. (\ref{lhos}) become zeros when $\phi=0$ and $\pi$. With using L'Hospital's rule, we have
\begin{equation}
\frac{d\phi}{d\varphi}=\frac{d(\mathcal{A}_1sin\varphi)/d\phi}{dL_{\phi}/d\phi}=\frac{d(\mathcal{A}_1sin\varphi)/d\varphi}{dL_{\phi}/d\phi}/(\frac{d\phi}{d\varphi}).
\end{equation}
\begin{equation}
\frac{d\phi}{d\varphi}=(\mathcal{A}_1\cos\varphi/\frac{dL_\phi}{d\phi})^{\frac{1}{2}}.
\label{eqn12}
\end{equation}

With the variable transformation, the scattering amplitude of Eq. (\ref{eq_phi}) turns to be\par

\begin{equation}
M_{\vec{p}_f}\approx \int_0^{2\pi}d\varphi C(\varphi)
e^{i(\mathcal{A}_0+\mathcal{A}_1\cos\varphi)}.
\label{eq_cphiappend}
\end{equation}
Where $C(\varphi)$ is the prefactor in the scattering amplitude integral which takes the following form
\begin{equation}
C(\varphi)=e^{i\mathcal{A}_0}\mathcal{F}_s\mathcal{D}_s p_{\rho_{s}}(\frac{\partial^2\mathcal{A}_s}{\partial t_s^2})^{-\frac{1}{2}}(\frac{(2\pi i)^3\mathcal{A}_1\cos\varphi}{\frac{dL_\phi}{d\phi}\det(\frac{\partial^2\mathcal{A}_s}{\partial({p}_{zs},p_{\rho s})})})^{\frac{1}{2}}.
\label{eqna5}
\end{equation}
$\frac{dL_\phi}{d\phi}=-\frac{d^2\mathcal{A}_s}{d\phi_0^2}$. When $\varphi=0$ and $\varphi=\pi$, we can prove the property that (proved in Appendix B)
\begin{equation}
\frac{d^2\mathcal{A}_s}{d\phi_0^2}\det\frac{\partial^2\mathcal{A}_s}{\partial(p_{\rho s},p_{zs})}=\det(\frac{\partial^2\mathcal{A}_s}{\partial(p_{zs},p_{\rho s},\phi_0)}).
\label{eqn15}
\end{equation}

Furthermore,
\begin{equation}
   p_{\rho s}^2/\det(\frac{\partial^2\mathcal{A}_s}{\partial (p_{zs},p_{\rho s},\phi_0)})=(\det(\frac{\partial^2\mathcal{A}_s}{\partial \vec{p}_s^2}))^{-1}
\end{equation}

In semiclassical model, the prefactor $\mathcal{F}_s\propto(\det(\frac{\partial^2\mathcal{A}_s}{\partial \vec{p}_{s}\partial \vec{p}_{f}}))^{\frac{1}{2}}$. We can combine the two determinants in Eq. (\ref{eqna5}), $\det(\frac{\partial^2\mathcal{A}_s}{\partial \vec{p}_{s}\partial \vec{p}_{f}}))^{\frac{1}{2}}(\det(\frac{\partial^2\mathcal{A}_s}{\partial \vec{p}_s^2}))^{-1}=\det(\frac{\partial \vec{p}_s}{\partial \vec{p}_{f}})$.

Under the cylindrical symmetry of linearly polarized fields, we have
\begin{equation}
\det(\frac{\partial \vec{p}_s}{\partial \vec{p}_{f}})=\frac{p_{\rho s}}{p_{\rho f}}\det(\frac{\partial (p_{zs},p_{\rho s})}{\partial(p_{zf},p_{\rho f})})
\label{eqna8}
\end{equation}
Substitute Eq. (\ref{eqna8}) into Eq. (\ref{eqna5}), we get the prefactor of following form,
\begin{equation}
C(\varphi)\propto\frac{\mathcal{D}_s}{\sqrt{\frac{d^2\mathcal{A}_s}{dt_{s}^2}}}(\mathcal{A}_1\cos\varphi)^{\frac{1}{2}}\frac{p_{\rho s}}{p_{\rho f}}\det(\frac{\partial (p_{zs},p_{\rho s})}{\partial(p_{zf},p_{\rho f})}).
\end{equation}
Based on the above derivation, we can approximate the scattering amplitude integral in cylindrical coordinates to the form expressed by Bessel functions, and finally get the  Eq. (\ref{final}).

\section{}
Under the saddle point condition of our derivation,
\begin{equation}
  \frac{\partial \mathcal{A}_s}{\partial p_{z0}}=0\qquad\frac{\partial \mathcal{A}_s}{\partial p_{\rho 0}}=0
\end{equation}

Differentiating the above equation,
\begin{equation}
\begin{split}
d(\frac{\partial \mathcal{A}_s}{\partial p_{z0}})=\frac{\partial^2\mathcal{A}_s}{\partial p_{z0}^2}dp_{z0}+\frac{\partial^2\mathcal{A}_s}{\partial p_{z0}\partial p_{\rho 0}}dp_{\rho 0}+\frac{\partial^2\mathcal{A}_s}{\partial p_{z0}\partial \phi_0}d\phi_0=0\\
d(\frac{\partial \mathcal{A}_s}{\partial p_{\rho 0}})=\frac{\partial^2\mathcal{A}_s}{\partial pp_{\rho 0}^2}dp_{\rho 0}+\frac{\partial^2\mathcal{A}_s}{\partial p_{z0}\partial p_{\rho 0}}dp_{z0}+\frac{\partial^2\mathcal{A}_s}{\partial p_{\rho 0}\partial \phi_0}d\phi_0=0.
\end{split}
\end{equation}
According to the total differential formula, we have the following properties,
\begin{equation}
\begin{split}
  &d^2\mathcal{A}_s=\frac{\partial^2\mathcal{A}_s}{\partial p_{z0}^2}dp_{z0}^2+\frac{\partial^2\mathcal{A}_s}{\partial p_{\rho 0}^2}dpp_{\rho 0}^2+\frac{\partial^2\mathcal{A}_s}{\partial \phi_0^2}d\phi_0^2\\
  &+2\frac{\partial^2 \mathcal{A}_s}{\partial p_{z0}\partial p_{\rho 0}}dp_{z0}dp_{\rho 0}+2\frac{\partial^2 \mathcal{A}_s}{\partial p_{\rho 0}\partial \phi_0}dp_{\rho 0} d\phi_0+2\frac{\partial^2 \mathcal{A}_s}{\partial \phi_0\partial p_{z0}}d\phi_0 dp_{z0}
  \end{split}
  \label{b1}
\end{equation}
Use the above properties we can get the property we need in Appendix A.
\begin{equation}
\begin{split}
&\det(\frac{\partial^2 \mathcal{A}_s}{\partial(p_{z0},p_{\rho 0})})\frac{d^2\mathcal{A}_s}{d\phi_0^2}=(\frac{\partial^2\mathcal{A}_s}{\partial p_{z0}^2}\frac{\partial^2\mathcal{A}_s}{\partial p_{\rho 0}^2}-(\frac{\partial^2\mathcal{A}_s}{\partial p_{z0}\partial p_{\rho 0}})^2)\\
&\cdot(\frac{\partial^2 \mathcal{A}_s}{\partial \phi_0^2}+\frac{\partial^2 \mathcal{A}_s}{\partial p_{\rho 0}\partial \phi_0}\frac{dp_{\rho 0}}{d\phi_0}+\frac{\partial^2 \mathcal{A}_s}{\partial \phi_0\partial p_{z0}}\frac{dp_{z0}}{d\phi_0})\\
=&\frac{\partial^2\mathcal{A}_s}{\partial p_{z0}^2}\frac{\partial^2\mathcal{A}_s}{\partial p_{\rho 0}^2}\frac{\partial^2\mathcal{A}_s}{\partial \phi_0^2}-(\frac{\partial^2\mathcal{A}_s}{\partial p_{z0}\partial p_{\rho 0}})^2\frac{\partial^2\mathcal{A}_s}{\partial \phi_0^2}\\
&-(\frac{\partial^2\mathcal{A}_s}{\partial p_{z0}\partial \phi_0})^2\frac{\partial^2\mathcal{A}_s}{\partial p_{\rho 0}^2}-(\frac{\partial^2\mathcal{A}_s}{\partial p_{\rho 0}\partial \phi_0})^2\frac{\partial^2\mathcal{A}_s}{\partial p_{z0}^2}\\
&+2\frac{\partial^2\mathcal{A}_s}{\partial p_{\rho 0}\partial \phi_0}\frac{\partial^2\mathcal{A}_s}{\partial p_{z0}\partial p_{\rho 0}}\frac{\partial^2\mathcal{A}_s}{\partial p_{z0}\partial \phi_0}\\
=&\det(\frac{\partial^2\mathcal{A}_s}{\partial(p_{z0},p_{\rho 0},\phi_0)}).
\end{split}
\end{equation}

\end{document}